\documentclass[aps,pre,showpacs,twocolumn,superscriptaddress,floatfix]{revtex4-1}
\usepackage{graphicx}
\usepackage{color}
\usepackage{amsfonts}
\usepackage{amssymb,amsmath}
\usepackage{grffile}
\usepackage{pgfplots}

\begin{document}

\title{Role of the range of the interactions in thermal conduction}

\author{Carlos Olivares}
\affiliation{Department of Physics, PUC-Rio, Rio de Janeiro, Brazil}

\author{Celia Anteneodo}
\affiliation{Department of Physics, PUC-Rio, Rio de Janeiro, Brazil} 
\affiliation{Institute of Science and Technology for Complex Systems, Rio de Janeiro, Brazil}

\date{\today}

\begin{abstract} 
We investigate thermal transport along a one-dimensional lattice of classical inertial rotators, with attractive couplings 
which decrease with distance as $r^{-\alpha}$ ($\alpha \ge 0$), subject at its ends to Brownian heat reservoirs 
at different temperatures with average value $T$. 
By means of numerical integration of the equations of motion, 
we show the effects of the range of the interactions in the temperature profile and energy transport,  
and determine the domain of validity of Fourier's law in this context. 
We find that Fourier's law, as signaled by a finite  $\kappa$ in the thermodynamic  limit, 
holds only for sufficiently short range interactions, with $\alpha >\alpha_c(T)$. 
For  $\alpha<\alpha_c(T)$, a kind of insulator  behavior emerges at any $T$.

\end{abstract}

\maketitle

\section{Introduction}

Heat conduction is a hot topic in non-equilibrium physics~\cite{LepriBook2016,
Landi2013,LiLiuLiHanggiLi2015,WangXuZhao2015,DasDharSaitoMendlSpohn2014,MendlSpohn,
SavinKosevich2014, LiuLiuHanggiWuLi2014,XiongZhangZhao2014,YangZhangLi2010,WangWang2011,WangHuLi2012,LiRenWangZhangHanggiLi2012RMP}. 
While there is a large body of works contributing to understand the empirically observed Fourier's law of conduction,   
many issues are still challenging, specially in low dimensions 
(see \cite{ReviewBonettoLebowitzReyBellet2000,ReviewLepriLiviPoliti2003,ReviewDhar2008} and references therein). 
For one-dimensional systems, Fourier's law takes the  form $J=-\kappa \, dT/dx$, where 
$J$ is the flux, $dT/dx$ the temperature gradient, and $\kappa$ the heat conductivity, 
that depends on the system and can also depend on the temperature but not on 
system size. 
While, for systems with momentum non-conservative thermal noise~\cite{Landi2014,Dhar2011}, 
 an-harmonic pinned systems~\cite{Aoki2000}, or systems with local reservoirs~\cite{Bonetto2004}, 
normal transport (hence, finite $\kappa$) is observed. 
Differently, anomalous transport  typically occurs in other one-dimensional model 
systems~\cite{LiLiuLiHanggiLi2015,LiuHanggiLiRenLi2014,LiWang2003,NarayanRamaswamy2002,AokiKusnezov2001,ProsenCampbell2000,Nakazawa1970,RiederLebowitzLieb1967}. 
In those cases, the conductivity can exhibit a divergent dependence on system size  
and super-diffusion occurs~\cite{basile2009}, hence the Fourier's law is not satisfied.
This scenario  has been attributed to momentum conservation~\cite{ProsenCampbell2000}.  
In apparent contradiction,  the Fourier's law does hold for the momentum 
conservative model of rotators with nearest-neighbors 
interactions~\cite{Giardina1999, Gendelman2000, LiLiLi2015,YangHuComment2005,GendelmanSavinReply2005,YangGrassberger2003,PereiraFalcao2006}. 
But, in this case, another quantity, the  stretch,  is not conserved~\cite{DasDhar2015}. 
However, when extra mechanical forces act at the extremities~\cite{Iacobucci2011}, 
counter-intuitive effects on the flux and on the temperature profile  occur also in the rotators model. 

Besides that, it is known that the range of the interactions can bring  new features to a
system. 
Ensemble inequivalence, phase transitions, relaxation times that increase with systems size and 
formation of quasi-stationary states, amongst others, can emerge when the interactions are sufficiently 
long-range~\cite{Antoni1995, Moyano2006,Anteneodo1998,TamaritAnteneodo2000,RuffoReview2009,Antunes2015,
CampaGuptaRuffo2015,TelesGuptaCintioCasetti2015,GuptaDauxoisRuffo2016,MiloshevichNguenangDauxoisKhomerikiRuffo2015,
RochaFilhoSantanaAmatoFigueiredo2014,LevinPakterRizzatoTelesBenetti2014,BuylNinnoFanelliNardiniPatelliPiazzaYamaguchi2013,
NardiniGuptaRuffoDauxoisBouchet2012,GuptaMukamel2010}. 
In the present context, for instance, the long relaxation times observed in long-range systems~\cite{Moyano2006,TelesGuptaCintioCasetti2015, MiloshevichNguenangDauxoisKhomerikiRuffo2015}  might affect thermal conductivity.
Then, a natural question is: which is the influence of the range of the interactions  in heat conduction? 
Addressing this issue, which can bring new insights to the above scenario, is the aim of the present work.
For that purpose, we consider a paradigmatic model system governed by a Hamiltonian dynamics, which generalizes the rotators model. 
It is known in the literature as $\alpha$-XY~\cite{Anteneodo1998,TamaritAnteneodo2000,Anteneodo04}, 
whose parameter $\alpha$ allows to  adjust the  range of the interactions from the nearest-neighbors to the global (mean-field) cases.

\section{The system}

The model $\alpha$-XY consists of a chain of classical rotators, attached to the sites $i=1,\ldots,N$ of a one-dimensional lattice, 
whose  dynamics is governed by the   Hamiltonian~\cite{Anteneodo1998,TamaritAnteneodo2000} 
\begin{equation}
\begin{split}
  \mathcal{H}^\alpha &= \sum_{i=1}^N  \frac{L_i^2}{2 I_i}  + 
  \frac{\epsilon}{ 2 } \frac{1}{ \widetilde{N}^\alpha}  \sum_{i=1}^N  \sum_{j \neq i}^N \left[ \frac{1-\cos(\theta_i - \theta_j)}{r_{i,j}^\alpha}  \right] \\
                     &\equiv \sum_{i=1}^N \mathcal{H}^\alpha_i 
                     \equiv  \sum_{i=1}^N \left[ \mathcal{T}_i + \mathcal{U}^\alpha_i \right],
\end{split}
\label{hamiltonian}
\end{equation}
where $L_i$ is the angular momentum, $I_i$ the rotational inertia, $\theta_i$ the angular position of  
the classical rotor, and we set
\begin{equation}
 \widetilde{N}^\alpha=\frac{1}{N} \sum_i^N \widetilde{N}^\alpha_i,\;\;\; 
 \widetilde{N}^\alpha_i=\sum_{j \neq i}^{N} r_{i,j}^{-\alpha}, \;\;\;  r_{i,j} = |i-j| ,
\end{equation}
in order to get a proper Kac prescription factor~\cite{Kac1963}, that guarantees  extensive energies 
in the thermodynamic limit (TL) when $\alpha\le 1$  and  is  adequate to our (free) boundary conditions. 
In the limits $\alpha =0 $ and $\alpha \to \infty$, 
we recover the \emph{infinite range} or mean-field (m-f) and \emph{first nearest neighbors}  (n-n) cases, respectively.

At the two ends of the lattice we apply (short-range) heat baths with 
temperatures $T_L,T_R$, with $T_L>T_R$.
To model each  reservoir, we use a Langevin heat bath. Therefore, 
the equations of motion are 
 \begin{equation}
 \begin{split}
  \dot{\theta}_k &= \omega_k, \;\;\;\;\mbox{ for $k=1,2,\dots, N$}, \\
   I_k \dot{\omega}_k &= F^\alpha_{k}, \;\;\;\; \mbox{ for $k=2,3,\dots, N-1$}, \\ 
	I_1 \dot{\omega}_1 &= F^\alpha_{1}  - \gamma_L \omega_1 + \eta_L, \\
  I_N \dot{\omega}_N &= F^\alpha_{N}  - \gamma_R \omega_N + \eta_R, \\
 \end{split}
\label{eq:alphaXYEqMotion}
\end{equation}
where $\omega_i$ is the angular velocity,   
\begin{equation}
\begin{split}
 F^\alpha_l &= -\frac{\partial \mathcal{H}^\alpha}{\partial \theta_l}  =
 \sum_{k \neq l}^N  
 f^\alpha_{l,k}, \hspace{0.5cm}
   f^\alpha_{l,k} = 
 \frac{\epsilon}{\widetilde{N}^\alpha} 
 \frac{\sin(\theta_k - \theta_l)}{r^\alpha_{k,l}},
\end{split}
\end{equation}
$\gamma_{L/R}$ are the damping coefficients of the Langevin force and $\eta_{L/R}$ are white 
noises with correlations
\begin{equation}
\begin{split}
 \langle \eta_{L/R}(t) \eta_{L/R}(t+\tau) \rangle &= 2 \gamma_{L/R} T_{L/R} \delta(\tau),  \\
 \langle \eta_L(t) \eta_R(t+\tau) \rangle &= 0.
\end{split}
\end{equation}

The flux is defined through the energy continuity equation for each particle,  
$\frac{d}{d t}\mathcal{H}^\alpha_i = \sum_{k \neq i}^N J^\alpha_{i,k},$ and, under the 
condition of local stationarity,  
we obtain the  flow of heat towards the particle $l$ due to the particle $k$  
\begin{equation}
 J^\alpha_{l,k} = \frac{1}{2}  f^\alpha_{l,k} (\omega_k + \omega_l). 
\end{equation}
Hence, we define the flux from the left (right) particles towards the particle $l$ as
\begin{equation}
\begin{split}
  \mathcal{J}^{L,\alpha}_{l} \equiv  \langle \sum_{k < l}^N J^\alpha_{l,k} \rangle, \hspace{1cm}&
  \mathcal{J}^{R,\alpha}_{l} \equiv  \langle \sum_{k > l}^N J^\alpha_{l,k} \rangle .
\end{split}
\label{flux}
\end{equation}
In the stationary state $\mathcal{J}^{L,\alpha}_{l} = -\mathcal{J}^{R,\alpha}_{l}$, for all $l$.
Moreover, the ``temperature'' at each particle position is defined as twice the mean kinetic energy 
$ T_i = \langle I_i \omega_i^2 \rangle$, 
which allows to depict a temperature profile along the system length.


\section{Results}

 \begin{figure}[b!]
 \begin{center}
 \includegraphics[width=0.45\textwidth]{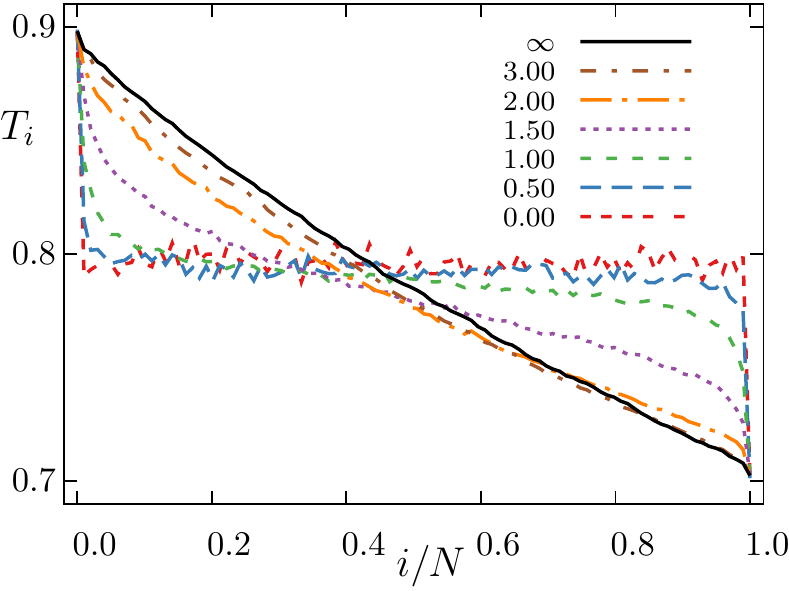}
\end{center}
 \caption{Temperature profiles for different values of $\alpha$ indicated on the figure. 
For each $\alpha$, averages over 50 realizations are computed. 
$T\equiv (T_L+T_R)/2 =0.8$, $\Delta T \equiv T_L-T_R= 0.2$, and  $N=100$.
}  
 \label{fig:profile}
\end{figure}

The equations of motion (\ref{eq:alphaXYEqMotion}) were integrated by means of a Brownian 
dynamics protocol~\cite{AllenTildesley1987,Paterlini1998}  
that reduces to a velocity-Verlet algorithm  in the absence of interactions with the heat reservoirs, 
as in the case of the bulk particles, which are not directly coupled to the reservoirs. 
The  fixed time step $dt$ for numerical integration was selected  so as to keep the energy of the 
corresponding isolated system constant within an error $\Delta E/E$ of order $10^{-4}$. 
%
Initial conditions  ($t = 0$) were set as follows:  angles and momenta were randomly chosen around zero 
(and the average momentum subtracted), within intervals adequate to reproduce the equilibrium temperature 
$T\equiv (T_L+T_R)/2$ for the isolated system, before switching on the reservoirs. 
After the thermal baths are connected and a transient has elapsed, 
the quantities of interest were averaged over 
at least 30 different initial conditions and along a time interval $\Delta t = 10^6$. 
Without loss of generality, in the numerical simulations reported here, 
we fixed the following values of the parameters of the Hamiltonian (\ref{hamiltonian}): 
$\epsilon=2$, $I_i=1.0$ for all $i$, and $\gamma_L=\gamma_R=1.0$.


In Fig.~\ref{fig:profile}, we show typical temperature profiles  for 
different values of $\alpha$.  
These profiles for $T=0.8$ and $\Delta T=0.2$ do not change substantially for values of $N$ larger than the value 
used in the Fig.~\ref{fig:profile}. 
We observe that, in the bulk region,  the profiles are almost linear. 
The absolute value of the slope is very close to its maximal value $\Delta T$   for n-n interactions ($\alpha \to \infty$), 
but the curves become less steep as $\alpha$ decreases. 
For sufficiently long-range  interactions, when $\alpha<\alpha_c \simeq 1$, the bulk profile becomes flat, and also more noisy.

\begin{figure}[b]
 \centering
 \includegraphics[width=0.5\textwidth]{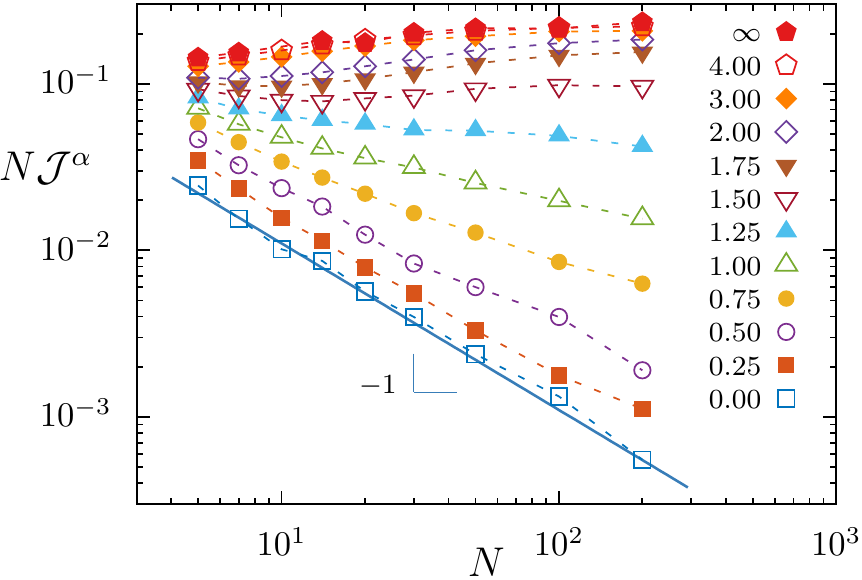}
 \caption{ 
Scaled flux $N\,\mathcal{J}$ vs $N$, for different values of $\alpha$ indicated on the figure. 
Dashed lines are guides to the eye. The relative standard error associated to each symbol is  about  100\%, 10\% and 1\% for 
 $\alpha \in$ $(0,1)$, $(1,2)$ and  $(2,\infty)$, respectively.
The solid line, with slope -1, was drawn for comparison.
$T\equiv (T_L+T_R)/2 =0.8$ and  $\Delta T \equiv T_L-T_R= 0.2$. 
}
 \label{fig:flux}
\end{figure}


The  stationary flux $\mathcal{J}^\alpha$ through the chain was computed by averaging  Eq.~(\ref{flux}) over the bulk particles, 
namely, $\mathcal{J}^\alpha \equiv \langle \mathcal{J}^\alpha_l \rangle_{bulk}$. 
The scaled  flux $N\mathcal{J}^\alpha$, for fixed $\alpha$, is depicted 
as a function of the size $N$ in Fig.~\ref{fig:flux}, for the same values of the end temperatures considered in Fig.~\ref{fig:profile}. 
For $\alpha$ above  $\alpha_c$, the scaled flux grows with $N$ attaining a finite value,  
like in the n-n limit~\cite{Giardina1999, Gendelman2000, LiLiLi2015}. 
Differently, for   any $\alpha$ below  $\alpha_c$, the scaled flux monotonically decays with $N$. 
Therefore,  a distinct behavior of  $N\mathcal{J}^\alpha$ vs $N$ emerges for short and long range interactions.
In the limit case $\alpha=0$, the scaled flux presents a neat decay as $1/N$,  indicating that the flux vanishes in the  TL. 
Apparently, a decay towards zero flux also occurs for any $\alpha<\alpha_c$.
It is noteworthy that, although the temperature profiles for small $\alpha$  are very similar to those observed 
for  chains of  identical masses with n-n harmonic interactions~\cite{RiederLebowitzLieb1967}, differently, 
 in the latter case  the flux is significantly non null.

Let us analyze the  m-f case ($\alpha=0$). 
In the TL, a null current along the chain is expected, because, 
on the one hand,    
each rotor interacts with each other rotor with equal intensity, 
on the other,
the contribution of  the end rotors 
(which are the only ones able to break the  m-f symmetry) 
becomes negligible compared to the interaction with the bulk. 
Therefore, the lattice structure where rotors are  attached is superfluous and there 
is not a preferential direction for the flux.  
From another viewpoint, it is  as if the two reservoirs were placed anywhere. 
Even if the current through the chain decays in the TL, 
there is a flow of energy from  the hot to the cold reservoir. 
In fact, the end rotors are directly coupled to the heat reservoirs and, as a consequence, 
there is a current from the hot bath to the first rotor, which is the same current from the last rotor  to the cold reservoir. 
This current is split in several paths: one is the short-circuit given by the long-range coupling between the two end particles and 
other paths go through  each rotor $i$, that is, passing from the first particle through the bulk particle $i$ towards the last 
one. But, noteworthily, a net current  does not pass ``through'' the chain, then  $N{\cal J}^\alpha=0$. 
For a finite system, even if the bulk rotors still interact globally, 
a small current  exists, because  in that case the 
effect over each bulk rotor due to the end rotors is not negligible. 
Consistently with this view, the current $N{\cal J}^\alpha$ decays with $N$.
This scenario which is clear for the m-f case ($\alpha=0$), apparently  also emerges for any $\alpha$ below $\alpha_c$,  
while, of course, it breaks down for sufficiently short-range interactions.

Once the bulk flux  due to given applied end temperatures  is computed,  
the heat conductivity $\kappa$ can be estimated through  
\begin{equation} \label{Jlin}
 \mathcal{J} \simeq  \kappa \Delta T/N, 
\end{equation}
for small enough difference between the temperatures applied at the ends,  $\Delta T$.  
Therefore, the plots   in Fig.~\ref{fig:flux} for $N\mathcal{J}^\alpha$  
vs $N$ (which were computed for fixed $T_L$ and $T_R$),  
directly reflect the behavior of the thermal conductivity  $\kappa$ 
with $N$ and $\alpha$. 
 
We can  observe  in Fig.~\ref{fig:flux} that, for fixed size $N$, the scaled flux (hence $\kappa$),  
continuously increases with $\alpha$. Therefore, short range interactions favor heat transport.
Above $\alpha \simeq 3$,  $\kappa$ practically attains the level of the n-n dynamics for all $N$.

For fixed $\alpha$, when the range is short enough ($\alpha>\alpha_c$),  
we notice that  $\kappa$ tends to a finite value in the large size limit, 
indicating the validity of  Fourier's law.   
Differently,  below $\alpha_c$,   $\kappa$ decreases with $N$, 
apparently following a power-law decay $N^{-\beta}$, 
where $\beta \simeq 1$ for $\alpha=0$,  and  the exponent $\beta$
decreases with $\alpha$, vanishing above  $\alpha_c$.
This suggests that $\kappa$ becomes null in the TL for systems 
with $\alpha<\alpha_c$. 
However, much larger sizes, which are computationally infeasible, 
would be required to determine the precise decay law.


\begin{figure}[b!]
 \centering
\includegraphics[width=0.5\textwidth]{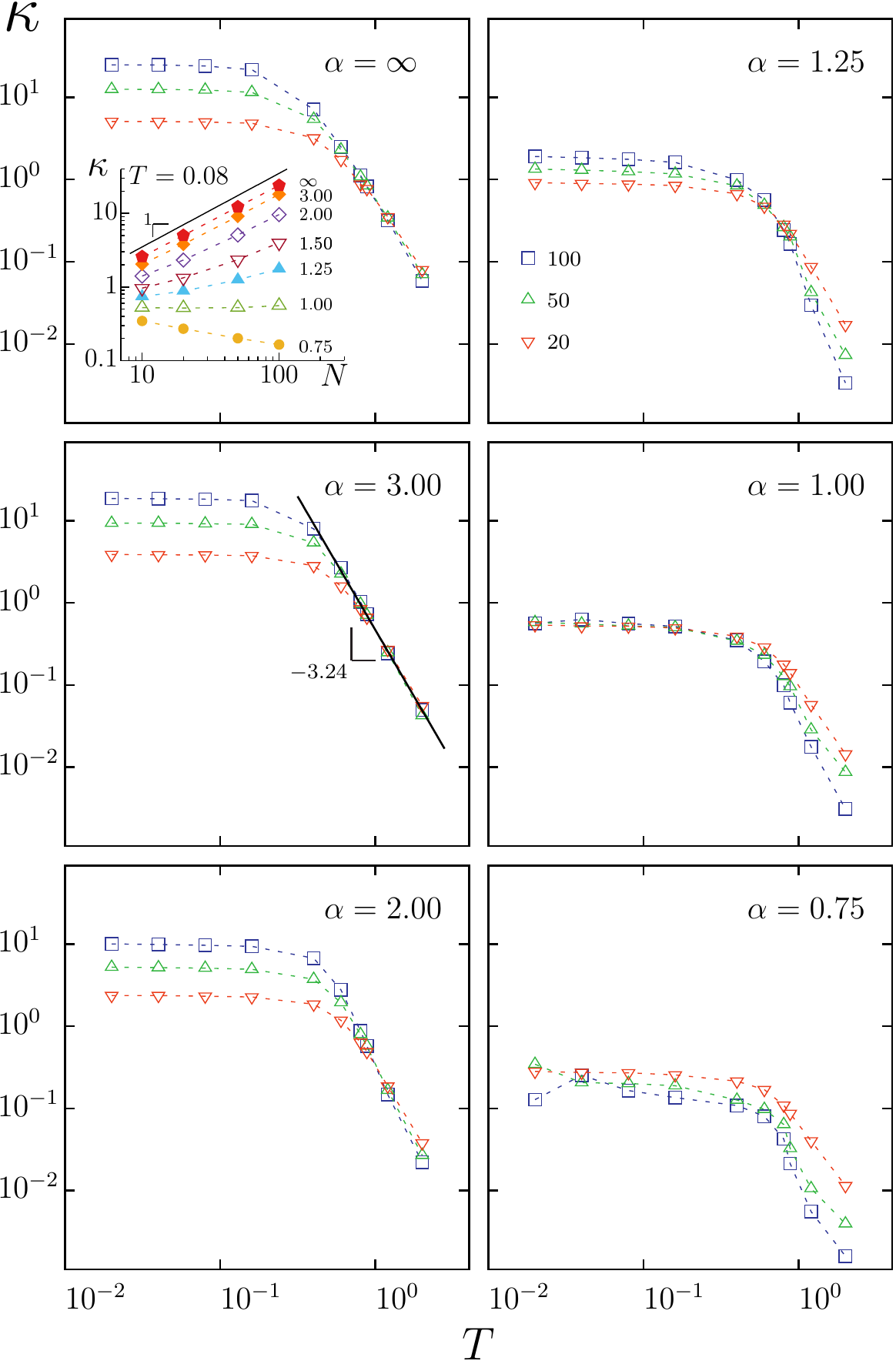}
 \caption{Conductivity $\kappa$ vs temperature $T$,  for different  system sizes and fixed $\alpha$ indicated on each panel.  
In case $\alpha=3$, we show a power-law fit to the plots of $\kappa(T)$ in the high temperature range. 
In the inset, we show the dependency of $\kappa$ with $N$ at a low temperature ($T=0.08$), for different values of $\alpha$.  
In all cases, $\Delta T/T=0.25$. Dotted lines are guides to the eye.   
}
 \label{fig:kappa}
\end{figure}

We also investigated the dependency of $\kappa$ with the mean temperature $T$, 
as depicted in  Fig.~\ref{fig:kappa}. 

Let us comment first on the dependency of $\kappa$ with  $T$ for fixed size $N$. 
For any $\alpha$, we observe that, 
 the heat conductivity   does not depend on $T$ at low temperatures,   but   
there is a crossover to a high temperature regime where $\kappa$ decays with $T$. 
In the limit of sufficiently low temperature,  
particles feel the nearly harmonic bottom of the  potential well, 
reaching the limit of harmonic oscillators.
At very high temperatures, a decay with $T$  is expected, because,  
the kinetic energy becomes much larger than the potential one (which is bounded for the $\alpha$-XY Hamiltonian), 
then, rotors tend to behave as independent particles,  and, concomitantly,  any transport process tends to become hindered.

Next we  discuss the impact of $\alpha$ on the dependency of $\kappa$ with $N$.
Let us look at  the three leftside panels of Fig.~\ref{fig:kappa}. 
The scenario for any $\alpha \ge 2$ is the same observed for  n-n interactions ($\alpha\to\infty$).
At low temperature, $\kappa$  increases with $N$  (see also the inset, for $T=0.08$).
The high temperature decay seems to follow a power-law, with the same exponent 
observed for  n-n interactions (about -3.2~\cite{LiLiLi2015}),  
as illustrated for $\alpha=3$ in Fig.~\ref{fig:kappa}. 
Furthermore, at high $T$,  the curves for different sizes tend to coincide, consistently with our previous 
observation of a limiting value of $\kappa$ in the large size limit, when we discussed Fig.~\ref{fig:flux} which 
was built for $T=0.8$.

When the interactions are short-range ($\alpha > 1 $), the level of the flat region of $\kappa(T)$, 
observed for low temperatures below the crossover,  grows linearly with $N$ (see inset in the upper left panel), 
as  expected for a chain of harmonic oscillators. 
However, notice that, concomitantly with that increase of the  flat  level with $N$, 
the crossover temperature diminishes with $N$,  
suggesting that the curves $\kappa(T)$ for different values of $N$ tend to adhere to a same curve as $N$ increases 
(which occurs progressively at lower temperature  and larger conductivity).
If that were the case, the growth of $\kappa$ with $N$ would persist only at null temperature, 
but, at a given finite $T$,  the conductivity would increase sublinearly with $N$ stabilizing at a finite value in the TL. 
Then, Fourier's  law would hold. 
This possibility is in accord with previous claims~\cite{YangHuComment2005,YangGrassberger2003,LiLiLi2015} 
against the  divergence of the conductivity in the TL~\cite{Gendelman2000,GendelmanSavinReply2005,PereiraFalcao2006}, for  
n-n interactions, but much higher sizes would be required to confirm the result for any $\alpha>1$.

At the particular value $\alpha=1$, we observe that the flat region of the 
conductivity profile coincides for different values of $N$ (constant $\kappa$ in inset plot), 
while for $\alpha < 1$ (illustrated by the case $\alpha=0.75$), 
 the conductivity decays with system size. 
This points that $\alpha=1$ is a marginal case at low temperatures.

At very low temperatures, a particle feels an harmonic potential, regardless the value of $\alpha$.
However, in the case of long-range interactions, the harmonic approximation does not come solely from n-n interactions. 
Then the level of the flat region does not increase with $N$, but differently to the harmonic chain behavior, 
$\kappa$ decreases with $N$, becoming presumably vanishingly small in the TL.

Now consider the rightside panels of Fig.~\ref{fig:kappa} (that is, $\alpha < 2$), as well as the inset plot. 
For temperatures above the crossover, a distinctive feature in comparison 
with the leftside panels appears:  
the conductivity decays with $N$, suggesting vanishingly small values in the TL 
(like in Fig.~\ref{fig:flux}). 

Therefore, for long-range interactions ($\alpha<1$), $\kappa$ decays with $N$ for any $T$, 
in accord with our discussion for the m-f case where the conductivity vanishes 
in the limit $N\to \infty$ at any finite temperature.

\begin{figure}[b!]
 \centering
 \includegraphics[width=0.55\textwidth]{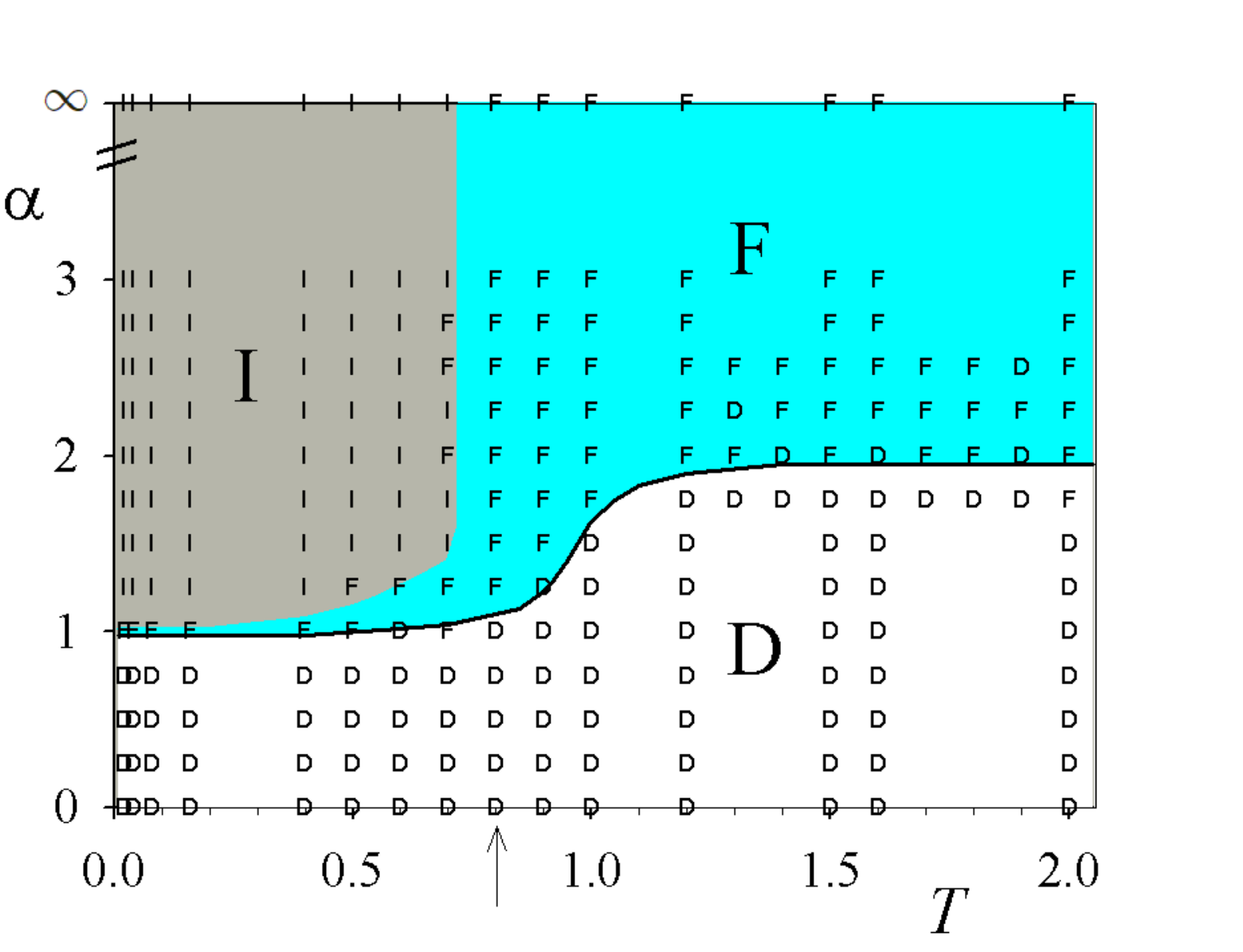}
 \caption{Diagram in the plane $T-\alpha$ of the different thermal regimes 
 of the conductivity  $\kappa$:  conductivity attains a finite value (F) or not, 
either decreasing (D) or increasing (I)  with $N$, for the investigated range of $N$. 
 From plots of $\kappa\sim N\mathcal{J}$ vs $N$, 
 like in Fig.~\ref{fig:flux}, we obtained the classification indicated by 
characters for each pair  ($T$,$\alpha$), with  $\Delta T/T=0.25$. 
The arrow  highlights the outcomes for  $T=0.8$,  in agreement with the 
results shown in Fig.~\ref{fig:flux}, although 
$\Delta T$ is slightly different. 
The black solid line represents the critical value $\alpha_c(T)$.
}
 \label{fig:diagram}
\end{figure}

The observed behavior of   $\kappa$ with $N$, for different values of $T$ and $\alpha$, 
is summarized  in the diagram of Fig.~\ref{fig:diagram}. We indicate whether, for the studied range of $N$,  
 $\kappa$ vs $N$  attains a finite value (F) or not, and in this latter case whether the behavior is 
decreasing (D) or increasing (I) with $N$. 
Much larger sizes would be required to determine the TL. 
This is  limited by the computational cost, taking into account that 
the integration algorithm is of order $N^2$, and  stabilization times also increase with $N$. 
Moreover, deviations from the diagram shown in  Fig.~\ref{fig:diagram} may occur for a different 
 ratio $\Delta T/T$, therefore this dependency should be also investigated, but again this is limited 
by the computational capacity.

Although one would expect a priori a threshold at $\alpha=d=1$ (where $d=1$ is the lattice dimensionality), 
the curve $\alpha_c(T)$ that emerges from simulations 
goes from  $\alpha_c\simeq 1$ at low temperatures to $\alpha_c\simeq 2$ at high temperatures. 
Also in this case, a distortion due to finite sizes can not be discarded.
But, coincidentally, the change of regime occurs near the critical temperature ($T^\star=1$) 
at which the isolated system in equilibrium suffers a ferro-paramagnetic transition when $\alpha\le 1$ 
(a transition also exists for $1<\alpha<2$, although at smaller $T^\star$)~\cite{TamaritAnteneodo2000}.

\section{Final remarks}

In summary, we have shown the portrait of heat conduction for a one-dimensional  system of interacting
particles as a function of the range of the interactions. 
The different domains are sketched in the diagram of Fig.~\ref{fig:diagram}.
We conclude that, the longer the range of the interactions, more the thermal conduction is spoiled. 
An interesting finding is the occurrence of a kind of insulator behavior for $\alpha<\alpha_c(T)$ 
(white region, denoted by `D', in the diagram of Fig.~\ref{fig:diagram}).  

For $\alpha >\alpha_c$, 
the thermal behavior is analogous to that found for nearest neighbor interactions. 
That is, for high temperatures, the thermal conductivity stabilizes at a finite value (`F') in the 
large size limit, decaying with $T$ following a power law. 
For low temperatures, the conductivity increases with $N$ (`I'). 
Although, for the studied sizes, we do not observe stabilization of $\kappa$  at a finite value, 
a picture similar to that observed for n-n interactions~\cite{YangHuComment2005,YangGrassberger2003,LiLiLi2015} emerges, 
pointing to the validity of Fourier's law.

Differently, when interaction are sufficiently long-range ($0 \le \alpha \le \alpha_c$), 
the scaled flux $N{\cal J}^\alpha$ (hence the conductivity) decays with $N$ for any $T$ (`D'),  
apparently vanishing in the TL, like is expected in the limiting mean-field case $\alpha=0$.
Then,  the bulk system  presents a  flat temperature profile and it behaves  like an insulator, in the sense 
that the flux through the chain is vanishingly small. 
This poor thermal conductivity is consistent with the slow relaxation to equilibrium 
observed in long-range systems, like in the Hamiltonian Mean-Field ($\alpha=0$)~\cite{Moyano2006,TelesGuptaCintioCasetti2015} 
or a modified Fermi-Pasta Ulam model~\cite{MiloshevichNguenangDauxoisKhomerikiRuffo2015}, 
where collisional effects act over times that increase with $N$, differently to short-range systems.  
The impact on thermal conductivity of other recent results about perturbation propagation, where 
no finite group velocity limits the spreading of perturbations and supersonic propagation 
occurs~\cite{perturbations} in long-range systems, also deserves investigation.  
Finally, note that it is plausible that the nontrivial scenario here reported for one-dimension 
can be extended to  an arbitrary dimension $d$, for $\alpha \mapsto \alpha/d$, which might also deserve 
a future extension of this work.


{\bf Acknowledgments:} We thank Brazilian agencies CNPq and Faperj for partial financial support. 


 \end{document}